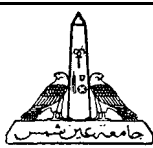

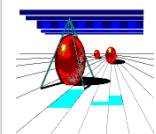

# Combined Algorithm for Data Mining using Association rules

**Walaa M.[1], Ahmed H.[2], Hoda K.[2]**

*1. School of Electronic Engineering, Canadian International College, Cairo campus of CBU*
*2. Ain Shams University, Faculty of Engineering, Computers & systems Department, Egypt*

| ARTICLE History | ABSTRACT |
|---|---|
|  | Association Rule mining is one of the most important fields in data mining and knowledge discovery. This paper proposes an algorithm that combines the simple association rules derived from basic Apriori Algorithm with the multiple minimum support using maximum constraints. The algorithm is implemented, and is compared to its predecessor algorithms using a novel proposed comparison algorithm. Results of applying the proposed algorithm show faster performance than other algorithms without scarifying the accuracy. |

**Keywords**

Association Rules; Minimum support; Apriori Algorithm

## 1. INTRODUCTION

It is increasingly important to develop powerful tools for analysis of the enormous data stored in databases and data warehouses, and mining interesting knowledge from it. Data mining is a process of inferring knowledge from such huge data.

Association rule mining searches for interesting relationships among items in a given data set. Association rule mining is used in many applications as economic and financial time series [4]. It is used in to identify software project success factors [14]. It is frequently used in Market Basket analysis [3], [7], [10].

For example, in a computer store there is a group of transactions and it is required to find what products are frequently bought together. Association rule can be represented as

$$Computers \Rightarrow Software \quad [Support=2\%, Confidence=60\%]$$

The rule means that customers, who buy computers, buy software as well. Rule support and confidence are two measures of rule interestingness; they reflect usefulness and certainty of discovered rules. Support of 2% means that 2% of all transactions contain both computers and software. Confidence of 60% means that 60% of the customers who buy computers, buy software as well.

Rules explosion that results from generating huge number of frequent itemsets especially in dense datasets is a problem of concern. Rule interestingness is a concept that is used to filter the useless and redundant rules as in [4], [12].

Some research work like [5], [13] discuss the rule generation problem, they suggested that mining Simple Association Rules (SAR) that have a single item as its consequent will be more efficient.

Given a single value to the minsup assumes that all items are of the same nature and have similar frequencies. Reference [6], [9], [11] dealt with multiple level items that represent hierarchies.

The paper is organized as follows: Section 2 explains mining association rules and describes several algorithms including Basic Apriori, SAR and multiple minimum supports using Maximum constraints. Section 3 presents the new algorithm and the novel comparison procedures which highlight the problem of comparing algorithms that use single minimum support value and other algorithms using multiple minimum support values. In Section 4, results of evaluating performance of the four algorithms are discussed. The conclusion is presented in Section 5.

## 2. MINING ASSOCIATION RULES ALGORITHMS

The problem of mining association rules is to generate all association rules that have support and confidence greater than the user-specified minimum support (called *minsup*) and minimum confidence (called *minconf*) respectively.

The problem of discovering all association rules can be decomposed into two sub problems:

(1) Finding all the frequent itemsets (whose support is greater than *minsup*), also called large itemsets.

(2) Generating the association rules derived from the frequent itemsets. If $X \cup Y$ and X are frequent itemsets, the rule $X \Rightarrow Y$ holds if the ratio of support $(X \cup Y)$ to support(X) is, at least, as large as *minconf*.

Since the solution to the second sub problem is straightforward [2], major research efforts have been spent on the first sub problem like [8], [9].

### 2.1 Apriori Algorithm

Apriori algorithm is an influential algorithm for mining frequent itemsets [1], [2]. The name of the algorithm is based on the fact that the algorithm uses prior knowledge of frequent itemsets properties.

Apriori employs an iterative approach known as a level-wise search, where k-itemsets are used to explore (k+1)-itemsets. First, the frequent 1-itemset is found, this is denoted by $L_1$, which is used to find the frequent 2-itemset $L_2$ and so on.

To improve the efficiency of the level-wise generation of frequent itemsets, a property called Apriori property is used to reduce the search space. This property states that all nonempty subset of a frequent itemset must also be frequent. A two step process is used to find $L_{k-1}$ from $L_k$

1) The join step: To find $L_k$, a set of k-itemsets is generated by joining $L_{k-1}$ with itself. This set of candidate itemsets is denoted $C_k$.

2) The prune step: $C_k$ is a superset of $L_k$, that is, its members may or may not be

frequent, but all the frequent k-itemsets are included in $C_k$. A scan of the database is done to determine the count of each candidate in $C_k$, those who satisfy the *minsup* is added to $L_k$. To reduce the number of candidates in $C_k$, the Apriori property is used. An example of Apriori algorithm is found in [2].

## 2.2 Mining Association Rules with multiple minimum supports using maximum constraints

Reference [9] proposed mining association rules with non-uniform minimum support values. This approach allowed users to specify different *minsup* to different items. They also defined the *minsup* value of an itemset as the lowest minimum supports among the items in the itemset. This is not always correct because it would consider some items that are not worth to be considered; just because one of the items in this itemset, its *minsup* was set too low. In some cases it makes sense that the *minsup* must be larger than the maximum of the minimum supports of the items contained in an itemset [8].

Reference [8] proposed an algorithm that gives items different minimum supports. The maximum constraint is adopted in finding frequent itemsets. That is, the *minsup* (denoted by *mI* for an itemset) is set as the maximum of the user specified minimum supports of the items contained in the itemset. Under the constraint, the characteristic of level-by-level processing is kept, such that the original Apriori algorithm can be easily extended to find the frequent itemsets. The algorithm first finds all the frequent 1-itemsets ($L_1$) for the given transactions by comparing the support of each item with its predefined *minsup*.

After that, candidate 2-itemsets $C_2$ can be formed from $L_1$. Note that, the supports of all the frequent 1-itemsets comprising each candidate 2-itemset must be larger than or equal to the maximum of their user specified *minsup*. This feature provides a good pruning effect before the database is scanned for finding large 2-itemsets.

The algorithm then finds all the large 2-itemsets $L_2$ for the given transactions by comparing the support of each candidate 2-itemset with the maximum of the user specified *minsup* of the items contained in the itemset. The same procedure is repeated until all frequent itemsets have been found. An example of the algorithm is found in [8].

## 2.3 Simple Association Rules

A simple rule is the rule with a single item as its consequent [5]. It is more efficient as the rule AB$\Rightarrow$C has the same meaning as A $\Rightarrow$BC.

It is proved that the rules that have multiple consequents can be derived from simple rules [5]. It has been observed that rule confidence (conf) with multiple items in its consequent could be represented by confidence of other rules each with a single item in its consequent. The following proof was given in [5].

$$\mathrm{conf}(A \Rightarrow BC) = \frac{\|ABC\|}{\|A\|}$$

$$= \frac{\|ABC\|}{\|AB\|} \times \frac{\|AB\|}{\|A\|}$$

$$= \mathrm{conf}(AB \Rightarrow C) \times \mathrm{conf}(A \Rightarrow B)$$

Thus, one may first concentrate on mining simple rules, based on which other rules concerned can be derived. Importantly, the set of simple rules is smaller in size than the original rule set but as 'equivalently' rich in semantics. An example of SAR is found in [5].

## 3. MINING SIMPLE ASSOCIATION RULES WITH MULTIPLE MINIMUM SUPPORTS USING MAXIMUM CONSTRAINTS

Based on the algorithms explained in section II, a new algorithm is proposed that mines simple association rules but with specifying different *minsup* to each individual item. The algorithm is called mining simple association rules with multiple minimum supports (SARMSMC).

### 3.1 The algorithm (SARMSMC)

The proposed algorithm is a combination between the two algorithms proposed in [5] and [8]. Figure 1 illustrates the algorithm steps which can be explained as follows:
1. A *minsup* is specified for each item, then check if each item's sup-count is greater than or equals its predefined *minsup* and generate the frequent 1-itemset in.
2. Candidate itemsets are generated only if each item's sup-count in this itemset is greater than or equal to the maximum predefined *minsup* specified for each of those items (*mI*).
3. Frequent itemsets are generated if itemset sup-count is greater than or equal to *mI*.
4. After finding the frequent k-itemsets, the itemset's subsets at level k-1 only are found then generate simple rules and check if the rule's conf is greater than or equals *minconf*.

In the example illustrated in Figure 2, the itemsets are generated according to the steps mentioned above; if the

*minconf* specified is 75%, seven rules are generated in Apriori, one of them is not a simple rule which is $A \Rightarrow BE$. In this example, this rule is not generated as it could be derived from the rules ($A \Rightarrow B$; $AB \Rightarrow E$; $A \Rightarrow E$; $AE \Rightarrow B$).

1. Create candidate one itemset $C_1$

2. Enter a *minsup* for each item in $C_1$

3. **for** each item in $C_1$

4.   **if** item sup-count >= its *minsup* **then** output $L_1$: item *//generate frequent 1- itemset*

5. Perform join as **Apriori**

6. **for** each itemset

7.   $mI$ = max of *minsup* of each item in itemset

8.   for each item in itemset

9.     **if** item sup-count >= *mI* **then** output $C_k$: itemset *// generate candidate itemsets*

10. **for** each itemset in $C_k$

11.   **if** itemset sup-count >= *mI* **then** output $L_k$: itemset  *// generate frequent itemsets*

12. **for** each $l_k$ in $L_k$  *// $l_k$ is a k-frequent itemsets of $L_k$*

13.   $SB = \{(k-1) - \text{itemsets } l^{k-1} \mid l^{k-1} \subset l^k\}$

14.     for each $l^{k-1} \in SB$

15.       conf = sup-count ($l^k_c$) / sup-count ($l^{k-1}$)

16.       **if** conf ≥ *minconf* **then** output $r$: $l^{k-1} \Rightarrow (l^k - l^{k-1})$  *// generate ruleset*

Fig. 1. SARMSMC Algorithm

### 3.2 Comparative Procedures between single and multiple supports algorithms

The comparison between any mining association rules algorithms is either made on number of rules generated by each algorithm or on their processing times. To compare the processing times, same algorithm parameters should be used which are the *minsup* and *minconf*. But using the same *minsup* causes confusion when comparing an algorithm that takes one *minsup* and other that takes multiple *minsup*. If the output of the single and multiple supports algorithms is the same, it means that both had equivalent parameters. The procedures illustrated in the flow chart of Figure 3 are used to specify a *minsup* to each item in order to unit the output of single and multiple supports algorithm. This will make comparing the processing times is based on a reliable aspect by uniting the output.

2. For each rule, determine which itemsets are contained in this rule.

3. Get the sup-count of each itemset.

4. Calculate the values of *mI* of each itemset.

5. For each item in the itemset, specify a *minsup* equals to *mI*.

6. If any item is specified more than one *minsup*, choose the smallest amount.

7. If some items were not specified any *minsup*, this means that they did not appear in the rules generated. They should be specified a *minsup* greater than their sup-count to be excluded from frequent 1- itemset.

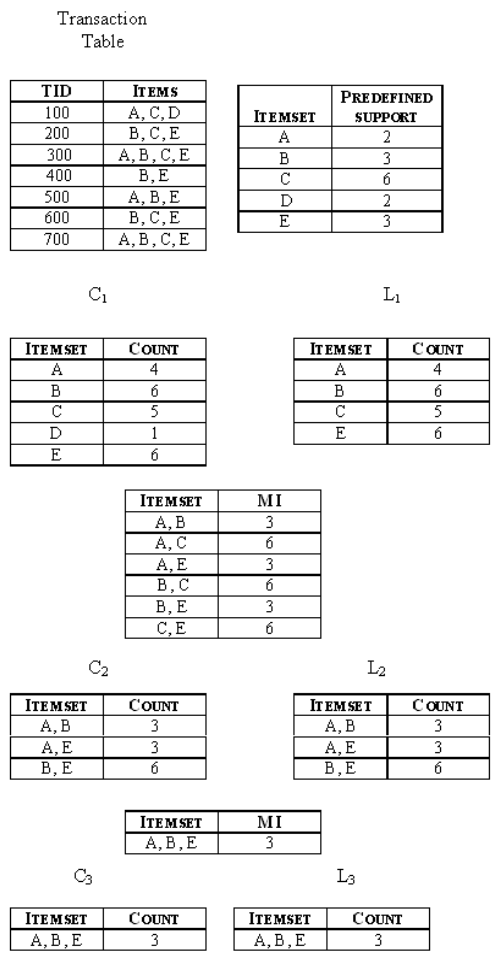

Fig. 2. SARMSMC Example

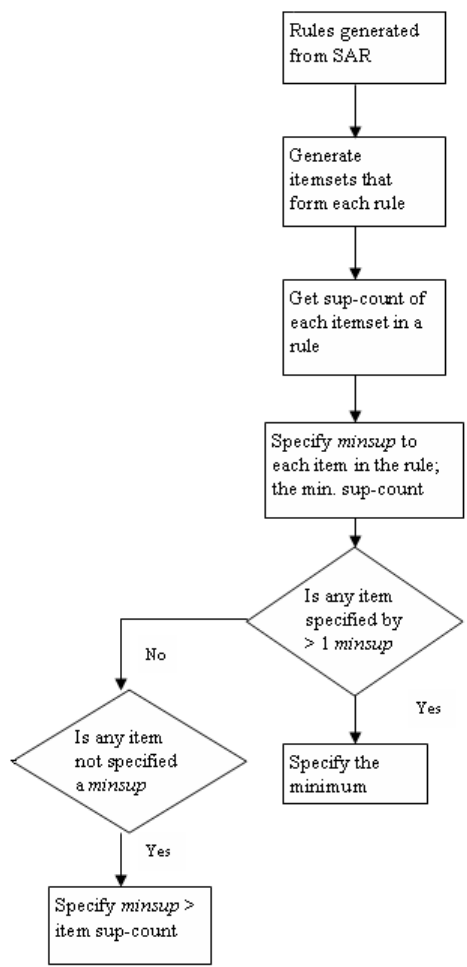

Fig. 3. Comparison Procedures

In Figure 3 these steps are taken to decide what *minsup* should be specified to each individual item when comparing SAR with SARMSMC.

The comparison procedures are:

1. Take the rules generated from SAR algorithm with a specific *minsup* i.e. 1%

For example:

After mining a group of transactions using SAR algorithm, a number of rules are generated.

If the first rule was:

AB⇒C

The itemsets in this rule and their sup-count are:

A, B      Sup = 589
A, C      Sup = 725
B, C      Sup = 1623
A, B, C   Sup = 589

The *mI* should not exceed any of the support counts specified above, If their specified *minsup* is equal to their least number = 589, we guarantee that those four itemsets are generated and so this rule will be generated.

minsup (A) = 589
minsup (B) = 589
minsup (C) = 589

If the second rule was:

DB⇒E

The itemsets in this rule and their support counts are:

D, B      Sup = 485
D, E      Sup = 559
B, E      Sup = 1513
D, B, E   Sup = 485

For those itemsets to be generated the *minsup* of those items should not exceed 485 which is the least number among their support count so,

minsup (D) = 485
minsup (B) = 485
minsup (E) = 485

But the minsup (B) was specified before to be equal to 589. In this case the smallest amount is chosen which is 485 to be sure that all itemsets that contain this item is generated.      Repeat the procedures mentioned above for every rule to find the *minsup* that should be specified to each item. The items which do not appear in the rules should be specified a *minsup* greater than its sup-count to be excluded from frequent 1-itemset.

## 4. PERFORMANCE EVALUATION

There are two datasets used to test the four algorithms. The first dataset is the real world dataset used in KDD CUP 2000 (http://www.ecn.purdue.edu/KDDCUP/) [15,16].

BMS-WebView-1 is a dataset contains several months' worth of clickstream data from two e-commerce web sites [14]. The second dataset is from AdventureWorksDW which is a sample database used frequently in SQL Server 2005. These databases are processed to be in the form of Transactions (TID, Items). After data processing and saving it in an XML file, the dataset generated from BMS-Web View-1 has the following characteristics:

   Number of transactions: 27,736
   Number of individual Items: 348
   Xml File size: 6.07 MB

The dataset generated from AdventureWorksDW has the following characteristics:

   Number of transactions: 21,255
   Number of individual Items: 37
   Xml File size: 5.58 MB

The comparison procedures explained in section 3 are applied when comparing Apriori with maximum constraints and when comparing SAR with SARMSMC.

### 4.1 Processing Time

The four algorithms mentioned in the paper are tested and evaluated for time and accuracy as illustrated in section B.

Follow the procedures mentioned in section 3, specify a *minsup* for each item to test the multiple supports algorithms and then generate rules at constant *minconf*. The processing time of each algorithm when applied on AdventureWorksDW is illustrated in Table 1 and the comparison graph is illustrated in Figure 4.

Table 1 Processing Times of Apriori, SAR, Maximum Constraints and SARMSMC for AdventureWorksDW

| Apriori Equiv. sup-count (%) | Apriori Time (sec) | SAR Time (sec) | Max. Constraints Time (sec) | SARM SMC Time (sec) |
|---|---|---|---|---|
| 0.2 | 97 | 94 | 93 | 91 |
| 0.175 | 109 | 105 | 104 | 101 |
| 0.15 | 119 | 113 | 112 | 109 |
| 0.125 | 134 | 126 | 127 | 122 |
| 0.1 | 156 | 142 | 148 | 138 |
| 0.075 | 196 | 169 | 181 | 164 |

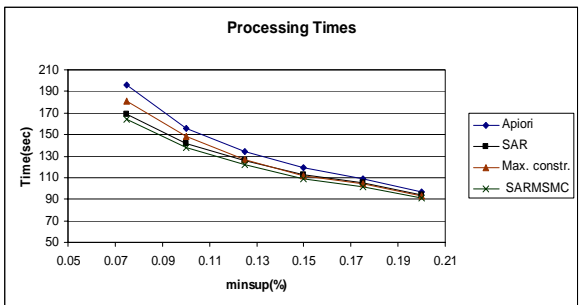

Fig. 4. Comparison of processing time between Basic Apriori, SAR, max. constraints and SARMSMC of AdventureWorksDW

In Figure 4 SARMSMC takes the least amount of time among other algorithms and Basic Apriori takes the longest time. The time increases when the *minsup* decreases because the number of itemsets and rules generated increase.

The processing time of each algorithm when applied on BMS-Web View-1 is illustrated in Table 2 and the comparison graph is illustrated in Figure 5.

Table 2 Processing Times of Apriori, SAR, Maximum Constraints and SARMSMC for BMS-Web View-1

| Apriori Equiv. sup-count (%) | Apriori Time (sec) | SAR Time (sec) | Max. Constraints Time (sec) | SARM SMC Time (sec) |
|---|---|---|---|---|
| 1 | 2741 | 2767 | 27 | 27 |
| 0.8 | 2742 | 2751 | 53 | 51 |
| 0.5 | 2750 | 2751 | 167 | 167 |
| 0.4 | 2760 | 2786 | 289 | 289 |
| 0.3 | 2787 | 2830 | 471 | 470 |

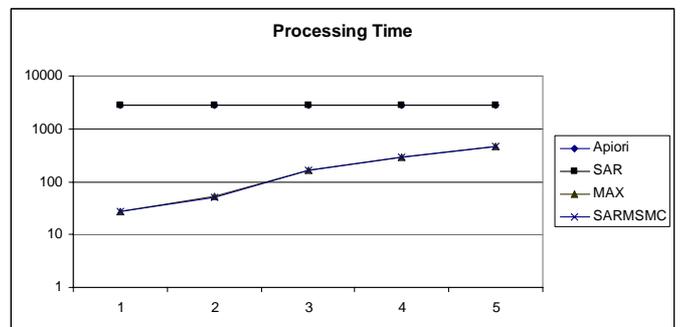

Fig. 5. Comparison of processing time between Basic Apriori, SAR, max. constraints and SARMSMC of BMS-Web View-1

In Figure 5 SARMSMC takes much less time than Apriori and SAR. The two single support algorithms almost consume the same time at different values of *minsup* and this time is bigger than the multiple supports algorithms. The time taken by multiple supports algorithm increase when the *minsup* decrease.

The main difference between the results of the two datasets used in comparison is the number of individual items. There are 348 items in BMS-Web View-1 opposing 37 items in AdventureWorksDW. That is the main reason that makes Apriori and SAR take very long time and almost the same processing time when applied on BMS-Web View-1 because of generating too many itemsets. The multiple supports algorithms are more efficient consuming much less time and generating the same number of rules when applied on this kind of dataset. There is no significant difference in time between generating all rules and simple rules on this kind of data because time used to generate itemsets are more significant.

Applying the four algorithms on the two datasets and after calculating the time taken to generate the itemsets and the time taken to generate the rules at constant *minsup* and *minconf* we get the following results.

Table 3 Itemset and Rule Generation Time

|  |  | Itemset Generation Time (sec) | Rule Generation Time (sec) |
|---|---|---|---|
| Adveunt ureWork sDW | Apriori | 137 | 12 |
|  | SAR | 137 | 3 |
|  | Max. constrai nts | 137 | 12 |
|  | SARM SMC | 129 | 3 |
| BMS-Web View-1 | Apriori | 2732 | 0.5 |
|  | SAR | 2740 | 0.4 |
|  | Max. constrai nts | 27 | 0.5 |
|  | SARM SMC | 26 | 0.4 |

Table 3 shows that generating the itemsets take much long time when the number of individual items are huge. Simple association rules only significantly differ in time than mining all rules when the number of individual items is small.

## 4.2 Accuracy Test, Interestingness Measurements and Algorithm Complexity

After building a mining model, the validity of the model should be tested. The data must be randomly separated into two separate datasets (training and testing). The training dataset is used to build the model, and the testing dataset is used to test the accuracy of the model. This is a part of the software engineering cycle to test many algorithms that solve the same problem then test their efficiency in solving the problem.

The two datasets are randomly separated to test the four algorithms mentioned in this paper. The separation was in the percentage of 10% and 90%. The four algorithms are applied on the training and testing datasets at different values of *minsup* with constant *minconf*. Follow the comparison procedures mentioned in section 3 to specify the *minsup* that should be given to each individual item while testing the multiple supports algorithm. The specification is a percentage of the *minsup* assigned to the original dataset, so the algorithms do not generate the same number of rules and the time is calculated as the time taken to generate each rule.

After performing the tests, the results are collected to calculate the accuracy of each algorithm by applying the concept of accuracy index where:

$$accuracy\,index = \frac{\Sigma accuracy}{\max(accuracy)}$$ (1)

Equation (1) shows that if an algorithm has the accuracy of 100%, it is the highest accuracy among the other algorithms. It does not mean that it has 100% absolute accuracy.

Applying the time index on the where:

$$time\,index = \frac{\Sigma time}{\max(time)}$$ (2)

Equation (2) shows that if an algorithm takes 100% time, it takes the longest amount of time among other algorithms.

Table 4 shows the results of the accuracy and time indices of AdventureWorksDW and Table 5 shows the results of the accuracy and time indices of BMS-Web View-1.

Table 4 Accuracy and Time indices of AdventureWorksDW

|  | Apriori | SAR | Max. Constr aints | SAM SMC |
|---|---|---|---|---|
| Time- inedx % | 100.00 | 97.77 | 83.45 | 81.40 |
| Accuracy -index % | 97.89 | 100.0 | 98.25 | 99.99 |

These results are illustrated in Figure 6 and Figure 7.

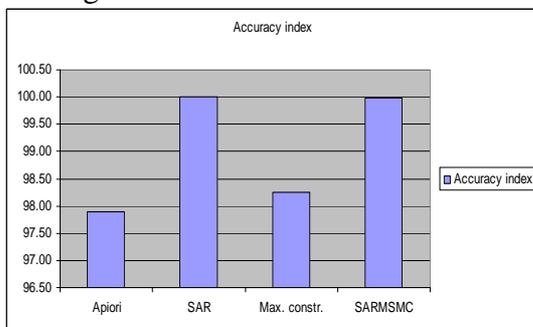
Fig. 6. Accuracy index of the four algorithms of AdventureWorksDW

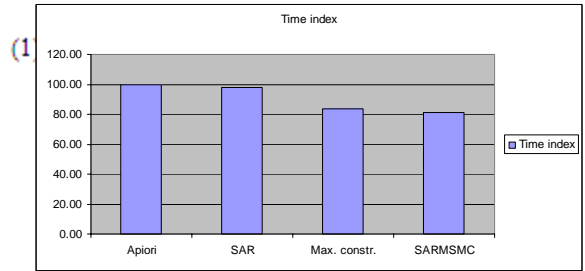
Fig. 7. Time index of the four algorithms of AdventureWorksDW

Figure 6 shows that SARMSMC is faster than other algorithms by 3%. Basic Apriori takes the longest time and SARMSMC takes the shortest. Figure 7 shows that the accuracy of the SAR is the best among other algorithms and the difference between the accuracy of SAR and SARMSMC is in the range of 0.01%.

Table 5 Accuracy and Time indices of BMS-Web View-1

|  | Apriori | SAR | Max. Constr aints | SAM SMC |
|---|---|---|---|---|
| Time- inedx % | 98.6 | 100 | 5.27 | 5.18 |
| Accuracy -index % | 99.13 | 100 | 92.05 | 85.92 |

These results are illustrated in Figure 8 and Figure 9.

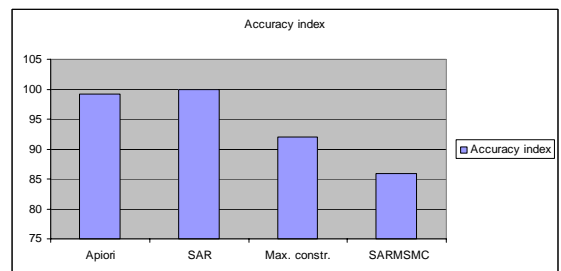
Fig. 8. Accuracy index of the four algorithms of BMS-Web View-1

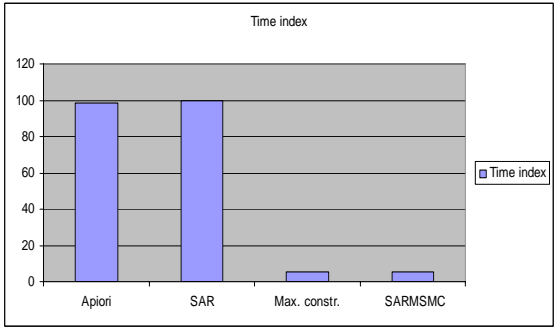

Fig. 9. Time index of the four algorithms of
BMS-Web View-1

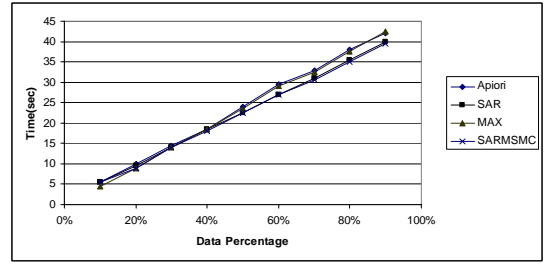

Fig. 10. Complexity of AdventureWorksDW

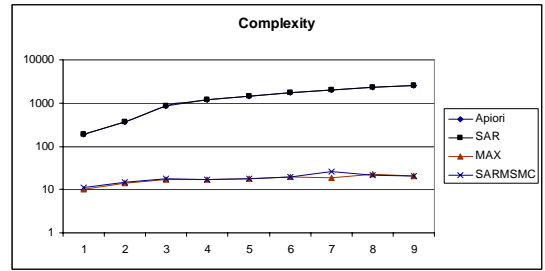

Fig. 11. Logarithmic graph shows the complexity of
AdventureWorksDW

Figure 8 shows that the accuracy of the SAR is the best among other algorithms and the differences in accuracy between SAR and SARMSMC is in the range of 15%. Figure 9 shows that SARMSMC is faster than other algorithms by 95%. It consumes almost the same time as Maximum constraints algorithm but much less time than Apriori and SAR.

The interestingness measurement of the rules could be calculated using the following equation [17].

$$lift(X \Rightarrow Y) = \frac{P(X \cup Y)}{P(X) \times P(Y)} = \frac{conf(X \Rightarrow Y)}{P(Y)} \qquad (3)$$

The rules generated from AdentueWorksDW and BMS-Web View-1 are 100% interesting because the support of the items has very low percentage. In BMS-Web View-1 the itemset support is in the range of 3% or 5%. In AdventureWorksDW, the maximum itemset support is in the range of 26%. So, the lift is all the time greater than one when using *minconf* equal to 50% or 30%.

Figure 10 and 11 are used to calculate the complexity of SARMSMC when applied on the two datasets. The algorithms are applied on different percentages of the datasets at const *minsup* and *minconf*.

In Figure 10, the time increases when the amount of data increases for the four algorithms. That means that SARMSMC has the same complexity as the other algorithms when applied on dataset that has few number of individual items.

In Figure 11, the time increases when the amount of data increases for Apriori and SAR. The time is almost constant when the amount of data increases for SARMSMC and Maximum constraints. That means that the complexity of SARMSMC is smaller than Apriori and SAR when applied on dataset that has huge number of individual items.

## 5. CONCLUSION

This paper presents an algorithm which is a combination between the simple associations rules derived from basic Apriori Algorithm with the multiple minimum support algorithms. The new algorithm is faster than any other algorithms and the accuracy is the best when applied on dataset that has little number of individual items. The new algorithm consumes almost the same time as the maximum constraint algorithm and the accuracy is affected in a bigger percentage when applied on dataset that has huge number of individual items. The rules generated from the algorithms are 100% interesting when applied on both datasets.

In this new algorithm the simple rules are generated and at the same time the user is given the flexibility to specify a different *minsup* for each individual item. This option overcomes the problem of rare items that need to be mined too. Generating simple rules decreases the processing time when applied on dataset that has little number of individual items. For other type of data, it affects the readability of the rules but does not affect the time.

A method to produce the exact same rules from multiple supports algorithms as generated by a single support algorithm is presented. It can be used to compare any single support algorithm with a multiple supports algorithm.

# 6. REFERENCES

1- Agrawal R., Imielinksi T., Swami A.. Mining association rules between sets of items in large database, in: The ACM SIGMOD Conference, 1993, pp. 207–216.

2- Agrawal R. and Srikant R. , Fast algorithms for mining association rules. In Proc. of the VLDB Conference, Santiago, Chile, September 1994. Expanded version available as IBM Research Report RJ9839, June 1994.

3- Agarwal, C., and Yu, P. Online generation of association Rule, ICDE-98, 1998, pp. 402-411.

4- Brin, S. Motwani, R. Ullman, J. and Tsur, S. Dynamic Itemset counting and implication rules for market basket data., SIGMOD-97, 1997, pp. 255-264.

5- Chen Guoqing, Qiang Wea, De Liu, Geert Wets. Simple association rules (SAR) and the SAR-based rule discovery @ 2002 Elsevier Science Ltd.

6- Han, J. and Fu, Y. "Discovery of multiple-level association rules from large databases." VLDB-95.

7- Lakshmanan, Ng. R. T., Han L. J. , Exploratory mining and pruning optimizations of constrained association rule., SIGMOD-98, 1998.

8- Lee, Yeong Chy, Hong, Tzung-Pei, Lin ,Wen-Yang. , Mining association rules with multiple minimum supports using maximum constraints @2005 Elsevier Inc.

9- Liu B.,, Hsu W., Ma Y.. Mining association rules with multiple minimum supports, in: The 1999 International Conference on Knowledge Discovery and Data Mining, 1999, pp. 337–341.

10- Park, J. S. Chen, M. S. and Yu, P. S., An effective hash based algorithm for mining association rules, SIGMOD-95, 1995, pp. 175-186.

11- Srikant, R., Agrawal, R., Mining Generalized Association Rules, Proceedings of the 21st VLDB Conference Zurich, Swizerland, 1995

12- Srikant, R., Vu, Q. and Agrawal, R. ,Mining association rules with item constraints, KDD-97, 1997, pp. 67-73.

13- Rastogi, R. and Shim, K. , Mining optimized association rules with categorical and numeric attributes, ICDE –1998.

14- Show-Jane Yen, Yue-Shi Lee. An efficient data mining approach for discovering interesting knowledge from customer transactions. @ 2005 Elsevier Ltd.

15- Girish K. Palshikar a, Mandar S. Kale , Manoj M. Apte. Association rules mining using heavy itemsets. @ 2006 Elsevier

16- Z. Zheng, R. Kohavi, L. Mason, Real world performance of association rule algorithms, in: Proc. 7th ACM SIGKDD Int. Conf. Knowledge Discovery and Data Mining (KDD-2001), 2001, pp. 401–406.

17- Agathe Merceron1 and Kalina Yacef. Interestingness Measures for Association Rules in Educational Data. [http://www.educationaldatamining.org/EDM2008/uploads/proc/6_Yacef_18.pdf]

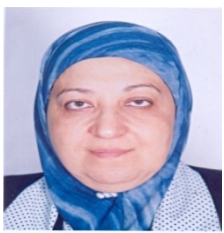

Dr. Hoda Korashy is an associative prof. at Ain shams university, Faculty of Engineering, computers & systems department. Field of interest: database and data mining, intelligent systems and agents.

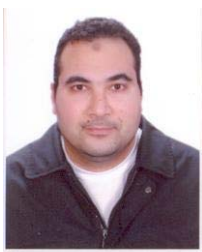

Dr. Ahmed Hassan is an assistant Professor in Computers and System Engineering department, Ain Shams University, Egypt. He is the vice director of ICTP government funded project in Egypt. Field of interest: software engineering, data mining and artificial intelligence applications.

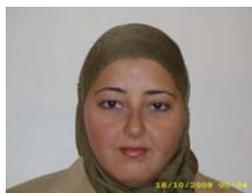

Walaa M. Medhat is an Engineering Instructor in School of Electronic Engineering, Canadian International College, Cairo campus of CBU. Field of interest: software engineering and data mining.

ملخص البحث

.

.

.

.

.